\documentclass[aps,prb,showpacs,twocolumn,longbibliography]{revtex4-1} %longbibliography adds titles to references

\usepackage[dvipdfmx]{graphicx}% Include figure files
\usepackage{dcolumn}% Align table columns on decimal point
\usepackage{bm}% bold math
\usepackage{hyperref}% add hypertext capabilities
\usepackage[utf8x]{inputenc}
\usepackage{pslatex}
\usepackage{textcomp} % for \textdegree
\usepackage{color}
\usepackage{lmodern} % nicer font - The Latin Modern fonts are enhanced versions of the Computer Modern fonts. They have enhanced metrics and glyph coverage.
\usepackage{amsmath}
\usepackage{amsfonts}
\usepackage{amssymb}

\graphicspath{{./figures/}}

\hypersetup{
    breaklinks = true,
    colorlinks = true,
    pdftitle = {}, 
    pdfauthor = {},
    pdfkeywords = {},
    linkcolor = blue,
    citecolor = blue,
    filecolor = black,
    urlcolor = blue
}

\newcommand{\yfeco}{Y(Fe$_{1-x}$Co$_x$)$_2$} 
 
\newcommand{\yfe}{YFe$_2$} 
\newcommand{\yco}{YCo$_2$}

\bibpunct{[}{]}{,}{n}{}{} %references in square brackets instead superscript 

\begin{document}
\begin{sloppypar}

\title{Electronic specific heat coefficient and magnetic properties of \\
 \yfeco{} Laves phases: \\
a combined experimental and first-principles study}

\author{Bartosz Wasilewski}
\author{Zbigniew Śniadecki}
\author{Miros\l{}aw Werwi\'nski}\email[Corresponding author: ]{werwinski@ifmpan.poznan.pl}
\affiliation{Institute of Molecular Physics Polish Academy of Sciences, ul. M. Smoluchowskiego 17, 60-179 Pozna\'{n}, Poland}

\author{Natalia Pierunek}
\affiliation{Faculty of Physics, Adam Mickiewicz University, Uniwersytetu Poznańskiego 2, 61-614 Poznań}
\affiliation{Institute of Molecular Physics Polish Academy of Sciences, ul. M. Smoluchowskiego 17, 60-179 Pozna\'{n}, Poland}

\author{J\'{a}n Rusz}
\author{Olle Eriksson}
\affiliation{Department of Physics and Astronomy, Uppsala University, Box 516, SE-751 20 Uppsala, Sweden}

\begin{abstract}
We investigated experimentally and computationally the concentration dependence of electronic specific heat coefficient $\gamma$ in \yfeco{} pseudobinary Laves phase system. 
The experimentally observed maximum in $\gamma$($x$) around the magnetic phase transition was interpreted within the local density approximation (LDA) combined with the virtual crystal approximation (VCA).
To explain the formation of the observed maximum, we analyzed theoretically the dependence of the magnetic energy, magnetic moments, densities of states, and Fermi surfaces on the Co concentration.
Furthermore, we carried out the calculations of density of states at the Fermi level as a function of fixed spin moment.
The calculated Co concentration at which $\gamma$ takes the maximum value ($x_{\textrm{max-LDA-VCA}}=0.91$) stays in good agreement with the measured value ($x_{\textrm{max-expt}} = 0.925$).
We conclude that the observed maximum in $\gamma(x)$ results from the presence of the sharp DOS peak in the vicinity of the Fermi level.

\end{abstract}

\date{\today}

\maketitle

\section{Introduction}\label{sec:intro}

The Laves phases are the largest group of intermetallic compounds~\cite{stein_structure_2004}.
They crystallize in close-packed structure classified into three types: hexagonal MgZn$_2$ (C14), cubic MgCu$_2$ (C15), and hexagonal MgNi$_2$ (C36)~\cite{murtaza_spin_2018,xie_new_2018,yan_structural_2018}.
%
%-----------------------mgcu2 - C15 - crystal structure---------------------------------
%
\begin{figure}[t]
\includegraphics[width=0.75\columnwidth]{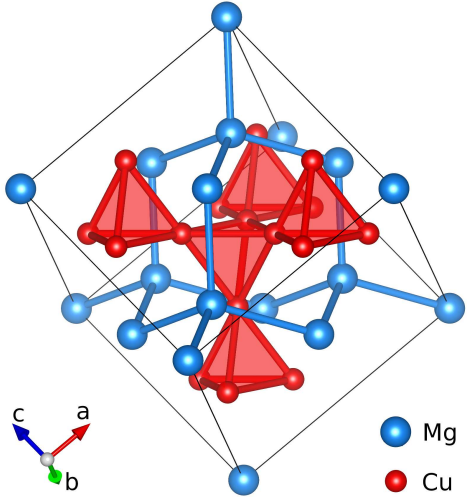}
\caption{\label{fig:struct}
Crystal structure of the cubic MgCu$_2$-type Laves phase. 
%
%The larger blue balls represent Mg atoms and the smaller red balls indicate Cu atoms. 
}
\end{figure}
\begin{table}[ht]
\caption{\label{tab:wyckoff} 
Atomic coordinates for \yfe{} and \yco{}, space group $Fd$-$3m$ (no. 227), origin choice two. 
}
\renewcommand{\arraystretch}{1.5}
\setlength{\tabcolsep}{1em}
\begin{tabular}{l|l|ccc}
\hline \hline
atom  	& site	& $x$ 	& $y$ & $z$ \\
\hline
Y   	& 8(a) 	& 1/8 	& 1/8   & 1/8 \\	
Fe/Co  	& 16(d)	& 1/2 	& 1/2   & 1/2 \\
\hline \hline
\end{tabular}
\end{table}
The \yfeco{} alloys crystallize in the cubic C15 MgCu$_2$-type structure~\cite{steiner_magnetische_1974} with the space group $Fd$-$3m$ (no. 227), see Fig.~\ref{fig:struct} and Table~\ref{tab:wyckoff}.
The primitive cell of \yco{} consists of two Y and four Co atoms.
The sublattice of Y has a diamond structure and the sublattice of Co forms the kagom\'{e} lattice (trihexagonal tiling).
%
%----------------yfe2, yco2, yfeco2 - magnetic properties ------------------
%
\yfe{} is a ferromagnet with the Curie temperature of about 550~K~\cite{piercy_evidence_1968,guzdek_electrical_2012},
while \yco{} is an exchange enhanced Pauli paramagnet~\cite{yamada_nmr_1980} undergoing a metamagnetic transition in a field of about 70~T (at 10~K)~\cite{schwarz_itinerant_1984,goto_itinerant_1990}.
Fe/Co alloying induces a paramagnetic--ferromagnetic phase transition in Y(Fe,Co)$_2$ system 
with a critical Fe concentration of about 0.14~\cite{kilcoyne_evolution_2000}.
The measured dependence of magnetic moment ($m$) on Co concentration starts at 2.80~$\mu_{\textrm{B}}$\,f.u.$^{-1}$ for \yfe{}, reaches a broad maximum for intermediate concentrations, and drops sharply to zero near the critical Co concentration~\cite{kilcoyne_evolution_2000}.
%
%----------------laves phases - general overview------------------
%
The pseudo-binary Laves phases \yfeco{} exhibit both extraordinary magnetic properties 
\cite{balamurugan_magnetism_2011, guzdek_electrical_2012, bonilla_formation_2014,kumar_permanent_2014,paul-boncour_metamagnetic_2019} 
and ability to absorb hydrogen~\cite{isnard_pressure-induced_2011,isnard_origin_2013,li_reversible_2016,paul-boncour_high_2017}.
Moreover, the DyFe$_2$/YFe$_2$ magnetic thin films are reversible exchange-spring magnets~\cite{stenning_exchange_2015,fu_quantitative_2016} and
the \yco{} alloys with rare-earth elements R$_{1-x}$Y$_{x}$Co$_2$ (R~=~Er,~Gd) are considered as magnetocaloric materials for application in magnetic refrigerators~\cite{baranov_butterflylike_2009,pierunek_normal_2017}.

%------------------previous results------------------
% 
Our previous experimental and theoretical investigations on the \yfeco{} system covered such issues as:
magnetic moments of \yfeco{}~\cite{eriksson_electronic_1988,eriksson_electronic_1989},
electronic structure of \yco{}~\cite{divis_electronic_2002,rusz_magnetism_2004},
effect of \yco{} doping~\cite{sniadecki_induced_2014}, 
magnetic percolation in \yfeco{} ~\cite{sniadecki_magnetic_2015}, 
Curie temperature of \yfeco{}~\cite{wasilewski_curie_2018}, and 
structural disorder in \yco{}~\cite{sniadecki_influence_2018}.

%---------------------electronic specific heat--------------------
%
In this work we focus on the concentration dependence of the electronic specific heat coefficient $\gamma$,
where the specific heat is a temperature derivative of the internal energy.
The temperature dependence of specific heat consists of lattice contribution and linear electronic term $\gamma\,T$.
As the Fermi-Dirac statistic indicates, only a small fraction of electrons contributes to the specific heat and the electronic contribution is the most pronounced in metals at low temperatures.
In the Sommerfeld model, the electronic specific heat coefficient $\gamma$ is calculated by converting the value of the density of states at the Fermi level ($\textrm{DOS}(E_{\textrm{F}})$) according to the equation 
\begin{equation} \label{eq1}
\gamma = \frac{1}{3} \pi^2 \textrm{k$_{\textrm{B}}^2$} \textrm{DOS}(E_{\textrm{F}}),
\end{equation}
where $\textrm{k$_{\textrm{B}}$}$ is the Boltzmann constant.

\section{Experimental and computational details}\label{exp_comp_details}

%----------------experimental details--------------
%
The ingots of several \yfeco{} compositions ($x$~=~0.85, 0.90, 0.925, 0.95, and 0.985) were prepared with the use of arc-furnace by repeated melting of required amounts of high purity Y (99.9\%), Co (99.9\%), and Fe (99.9\%) in Ar atmosphere.
The polycrystalline master alloys were rapidly quenched in a melt spinning device on a rotating copper wheel with the surface velocity of 40~m\,s$^{-1}$. 
The X-ray diffraction (XRD) with Co K$_{\alpha}$ radiation in Bragg–Brentano geometry was used to characterize the crystalline structure of the melt-spun flakes. 
Temperature dependences of heat capacity were measured with two-tau relaxation method using the Quantum Design Physical Property Measurement System (PPMS).

%------------------FPLO: supercells, CPA, VCA----------------------------
%
To simulate the chemical disorder in theoretical models of the considered \yfeco{} alloys three different methods were used: 
the coherent potential approximation (CPA)~\cite{koepernik_self-consistent_1997,soven_application_1970}, 
the virtual crystal approximation (VCA)~\cite{eriksson_electronic_1989}, 
and the ordered compound method (also called the supercell method)~\cite{eriksson_electronic_1988}.
The density functional theory (DFT) calculations were conducted with the full-potential local-orbital scheme (FPLO)~\cite{koepernik_full-potential_1999}. 
The older FPLO5.00-18 version of the code was used for the CPA, which is not available in the more recent versions. 
The rest of the calculations were performed with the FPLO14.0-49. 
For the exchange-correlation potential, we used the Perdew and Wang (PW92) model of the local-spin-density approximation (LSDA)~\cite{perdew_accurate_1992}.
As a result of using the CPA, we had to limit our calculations to the scalar-relativistic approximation.
It is because, that in the FPLO5.00-18 version of the code, the CPA calculation can only be carried out with scalar-relativistic approximation (not including spin-orbit coupling), therefore for consistency, this approach was also used in other cases.
The calculations were done using a $40\times 40\times 40$ \textbf{k}-mesh for the VCA and $16\times16\times16$ \textbf{k}-mesh for the CPA and ordered compound methods.
For all three approaches we used the criterion of simultaneous energy and density convergence with an accuracy of $\sim$2.72$\times$10$^{-7}$~eV (10$^{-8}$~Ha) and $10^{-6}$, respectively.
In the FPLO5 the Y$(4s, 4p)$ and Fe/Co$(3s, 3p)$ electrons were treated as semi-core with the Y~$4s$ and $4p$ orbitals having separate compression parameters, while the $3s$ and $3p$ orbitals in Fe and Co having a joint one.
The Y(5$s$, 5$p$, 4$d$) and Fe/Co(4$s$, 4$p$, 3$d$) electrons were treated as valence ones.
The considered crystallographic models were based on the optimized lattice parameters, 
as the application of the experimental parameters for \yco{} resulted in a ferromagnetic ground state, 
which is not consistent with the empirical data~\cite{rusz_magnetism_2004,khmelevskyi_magnetism_2005,sniadecki_influence_2018}.
%
%--------------optimized lattice parameters-----------------------
%
Due to the overbinding nature of the LDA, the calculated lattice parameters (7.05~\AA{} for \yfe{} and 6.95~\AA{} for \yco{}) are much smaller than the experimental ones (7.36~\AA{} for \yfe{} and 7.22~\AA{} for \yco{}~\cite{guzdek_electrical_2012}).
However, they stay in good agreement with the previous LDA results (7.04~\AA{} for \yfe{} and 6.96~\AA{} for \yco{}~\cite{zhang_collapse_2016,khmelevskyi_magnetism_2005}).
The lattice parameters for the intermediate \yfeco{} concentrations were interpolated assuming a linear behavior of $a(x)$ dependence, which is in a good agreement with experiment~\cite{guzdek_electrical_2012}. 
%
%---------------------ordered compound-----------------------------
%
In case of the ordered compound approach, we started with a model of \yfe{} supercell, composed of two primitive cells including four Y atoms and eight Fe atoms.
We consequently substituted Fe by Co in the Y$_4$Fe$_8$ master-cell producing a series of ordered ternary compounds: Y$_4$Fe$_7$Co$_1$, Y$_4$Fe$_6$Co$_2$, Y$_4$Fe$_5$Co$_3$, etc.
The detailed crystallographic data of these compounds are summarized in Table~\ref{tab:ordered_compounds} of the Appendix.
%
%--------------------vesta---------------------
%
VESTA code~\cite{momma_vesta_2008} was used for visualization of the crystal structure.

%------------gamma enhanced-------------------
%
We determined the enhanced specific heat coefficients $\gamma_{calc-enh}$ by multiplying the 
$\gamma$ values calculated from Eq.~\ref{eq1} by the so called enhancement factor ($\tilde{\gamma}$).
For each considered composition we used single value of enhancement factor $\tilde{\gamma}$ equal to 6.87,
which has been obtained by Tanaka and Harima for \yco{} by adjusting the $\gamma_{calc}$ to $\gamma_{expt}$~\cite{tanaka_mass_1998}.
Tanaka and Harima have introduced the enhancement factor
motivating that for the strongly correlated electron systems 
the many-body effects can be taken into account considered as self-energy of Co~$d$ electrons~\cite{tanaka_mass_1998}.
The enhancement factor ($\tilde{\gamma}$) was expressed as energy derivative of the self-energy ($\Sigma(\omega)$), 
\begin{equation} \label{eq2}
 \tilde{\gamma} = 1 - \{ \partial \Sigma(\omega) / \partial \omega \}_{\omega = E_{F}},
\end{equation}
where $\Sigma(\omega)$ was calculated with a method of second-order perturbation for Coulomb interactions ($U_{dd}$) between Co~$d$ electrons
in the framework of the Fermi liquid theory on the basis of a periodic Anderson model.
The value of $\tilde{\gamma} = 6.87$ has been obtainded by Tanaka and Harima by assuming $U_{dd}$ equal 1.8~eV~\cite{tanaka_mass_1998}.
Details of this method along with examples of applications can be found in Ref.~\cite{tanaka_calculations_1998}.

\section{Results and discussion}

\subsection{Experimental results and discussion}

\begin{figure}[ht]
\includegraphics[width=1.0\columnwidth]{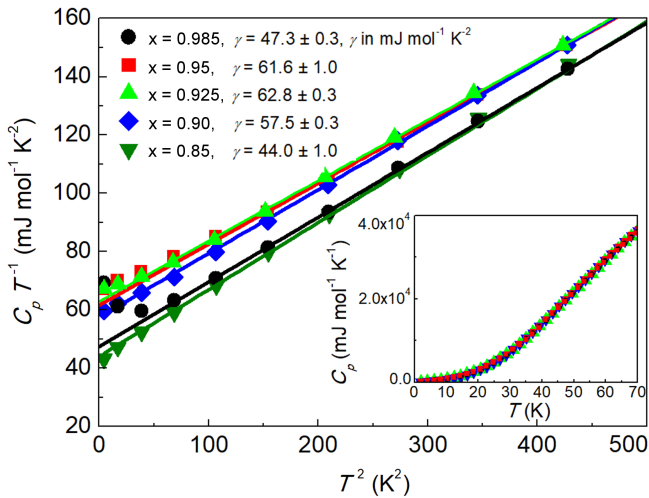}
\caption{\label{fig:cp_exp} 
The $C_p/T$ \textit{versus} $T^2$ dependences for several \yfeco{} compositions, where $x$~=~0.85, 0.90, 0.925, 0.95, and 0.985 (experiment -- symbols, fitting -- solid lines). 
Inset shows the temperature dependences of the specific heat ($C_p$).
}
\end{figure}

\begin{table}[ht]
\caption
{
\label{tab:gamma} 
The experimental electronic specific heat coefficient $\gamma_{\textrm{expt}}$, 
in mJ\,mol$^{-1}$\,K$^{-2}$, 
measured for \yfeco{} system in the Co-rich region ($0.85 \leq x \leq 0.985$). 
$\Delta$ is an estimated error of the measured value.
}
\renewcommand{\arraystretch}{1.5}
\setlength{\tabcolsep}{1em}
\begin{tabular}{c|ccccc}
\hline \hline
$x$ 
& 0.85 & 0.90 & 0.925 & 0.95 & 0.985 \\
\hline
$\gamma_{\textrm{expt}}$ 
& 44.0 & 57.5 & 62.8 & 61.6 & 47.3 \\
$\Delta$
& \textpm~1.0 & \textpm~0.3 & \textpm~0.3 & \textpm~1.0 & \textpm~0.3 \\
\hline \hline
\end{tabular}
\end{table}

\begin{figure}[ht]
\includegraphics[trim = 220 30 10 00,clip,height=1.05\columnwidth,angle=270]{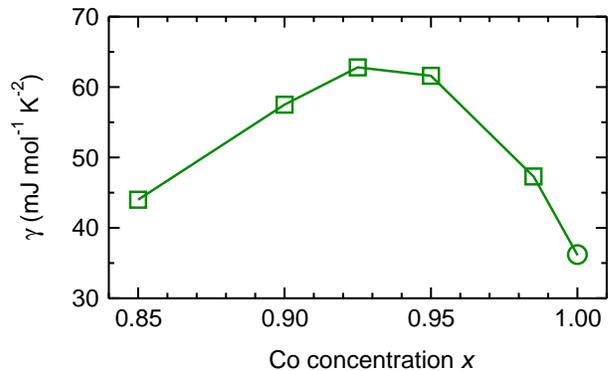}
\caption
{
\label{fig:gamma_exp} 
The experimental electronic specific heat coefficient $\gamma$ as measured for \yfeco{} system.
The value of $\gamma$ for \yco{} ($x=1.0$) comes from Muraoka et al.~\cite{muraoka_electronic_1977}.
}
\end{figure}

%-----------------specific heat measurements----------------
%
The experimental study of $\gamma(x)$ is conducted in the vicinity of Co concentration $x = 0.9$, where the maximum in $\gamma$ has been previously observed~\cite{muraoka_electronic_1977}.
However, the foregoing study has been carried out with a relatively large step of $\Delta x$~=~0.1.
Our results of specific heat ($C_p$) measurements in a temperature range between 2 and 70~K are presented on the inset of Fig.~\ref{fig:cp_exp}.
The $C_p(T)$ curves do not show any apparent differences and no indication of a long-range magnetic ordering is visible in the plots. 
However, the $C_p/T$ \textit{versus} $T^2$ dependences already reveal how the $\gamma$ coefficient changes with $x$.
The concentration dependence of the Sommerfeld coefficient exhibit a broad peak with maximum of 62.8~mJ\,mol$^{-1}$\,K$^{-2}$ at $x = 0.925$, see Fig.~\ref{fig:gamma_exp} and Table~\ref{tab:gamma}.
This result is in a good agreement with the previous one indicating the $\gamma_{max}$ slightly below 60~mJ\,mol$^{-1}$\,K$^{-2}$ located at $x = 0.9$ ~\cite{muraoka_electronic_1977}.

%-------------non linearity of the gamma results - magnetic percolation--------------------------
%
For $x$ equal to 0.925, 0.95, and  0.985, the estimations of $\gamma_{\textrm{expt}}$ were done excluding non-linear low-temperature regions.
Non-linearity, in the form of a significant upturn on the $C_p/T(T^2)$ dependence, is visible at low-temperatures for $x$~=~0.985 in the inset of Fig.~\ref{fig:cp_exp}. 
Some deviations from linearity were also detected for $x$~=~0.925 and 0.95. 
For $x$~=~0.85 and 0.90 the linear dependences in $C_p/T(T^2)$ were measured down to the lowest temperatures. 
A very small upturn of $C_p/T(T^2)$ for $x$~=~0.925 suggests that this alloy is just below the critical concentration for magnetic percolation. 
(Similar anomaly has been also observed for other Laves phases, as for Y$_{1-x}$Gd$_x$Co$_2$~\cite{baranov_onset_2003}.)
Thus, being careful one can determine the critical concentration as $0.90 < x_{crit} < 0.95$, 
with the other authors reporting somewhat lower values, $x_{crit} \sim 0.86$~\cite{kilcoyne_evolution_2000}, 0.88~\cite{steiner_magnetization_1979}, and 0.895~\cite{steiner_magnetization_1979}.
%
%------------magnetic clusters in melt spun samples---------------------------
%
Shift of the measured critical concentration can be explained as an effect of using the rapid quenching technique for synthesis of our alloys.
This results in chemical and topological disorder being introduced, which leads to the formation of magnetically ordered state~\cite{sniadecki_induced_2014,sniadecki_magnetic_2015,sniadecki_influence_2018}.
When starting from an exchange enhanced Pauli paramagnet \yco{}, an introduction of structural disorder (or addition of other element) causes the formation of magnetically ordered clusters (spin-glass-type behavior) with non-zero magnetic moment on Co atoms. 
Formation of such magnetic clusters has been described for Y$_{1-x}$Gd$_x$Co$_2$ as a 'microscopic' metamagnetic phase transition that occurs at a sufficiently high molecular field acting on Co atoms~\cite{baranov_onset_2003}. 
Magnetic percolation to the long-range magnetic ordering takes place when the volume and number of magnetic clusters, which can be described as localized spin density fluctuations, are increasing.
Such fluctuations originate from the distribution of $d$-$d$ exchange coupling due to the presence of chemical or structural disorder and from the inhomogeneous distribution of the local density of states. 
This picture leads to the conclusion that the additional contribution to the heat capacity of the samples above $x_{crit}$ is connected with the presence of magnetically ordered clusters.
The above suggests a close connection between the maximum in $\gamma(x)$ and magnetic phase transition in the considered Laves phases.

\subsection{Theoretical results and discussion}

According to Muraoka \textit{et~al.}~\cite{muraoka_electronic_1977} the characteristic enhancement of $\gamma(x)$ around the critical concentration should be attributed to the spin fluctuations.
Subsequently, two theoretical attempts have been made by Shimizu \textit{et~al.} to reproduce the maximum in $\gamma(x)$~\cite{terao_spontaneous_1983,yamada_magnetic_1987}.
The first approach was based on the Green's functions method~\cite{terao_spontaneous_1983} and 
in the second one the rigid band model was used on top of the \yco{} density of states (DOS) from the tight-binding approximation~\cite{yamada_magnetic_1987}.
Unfortunately, the $\gamma(x)$ dependences obtained from both methods are unsatisfactory, especially in the vicinity of the magnetic transition.

\subsubsection{Densities of states and electronic specific heat coefficient}
%
%-----------------------------dos------------------------------------------
%
%top  left 
\begin{figure}[t]
\includegraphics[trim = 30 30 80 35,clip,height=1.0\columnwidth,angle=270]{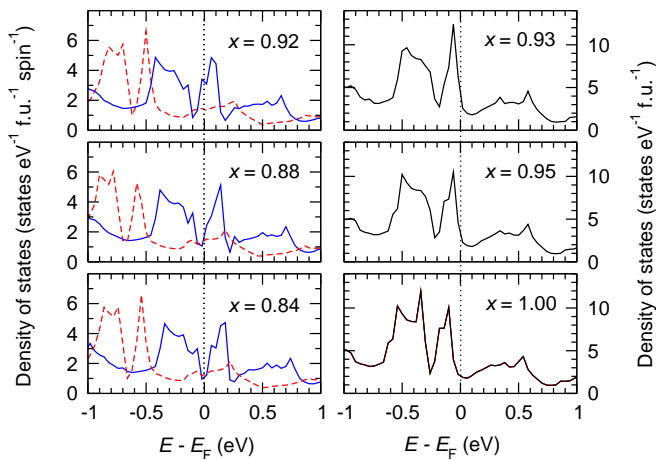}
\caption{\label{fig:dos}
To analyze the formation of maximum in $\gamma(x)$ on the Co-rich region of the \yfeco{} system, we employed the local density approximation (LDA) in combination with the virtual crystal approximation (VCA).
The calculated (with LDA-VCA) densities of states (DOS) for Co concentrations near $x_{crit} \sim 0.925$ at which a ferromagnetic--nonmagnetic phase transition occurs in \yfeco{} system.
Results of spin polarized calculations are presented on the left, with solid lines denoting majority and dashed lines denoting minority spin channels. 
The nonmagnetic results are shown on the right.
}
\end{figure}
In Fig.~\ref{fig:dos} we present the densities of states (DOS) for several Co-rich compositions of \yfeco{} system.
A comparison of total energies of ferromagnetic and nonmagnetic ground-state solutions indicates a magnetic phase transition at $x_{crit} \sim 0.925$. 
The presented DOS are spin polarized for a ferromagnetic region ($x < 0.925$) and nonmagnetic above the critical concentration ($x > 0.925$).
The valence bands of the considered alloys start at about -7~eV with the most significant contributions from the 3$d$ states located above -4~eV~\cite{tanaka_mass_1998,sniadecki_influence_2018}.
The DOS plots presented in Fig.~\ref{fig:dos} cover only the narrow region between -1 and 1~eV, 
which is the most important from the perspective of magnetic phase transformation.
For \yco{} we observe a characteristic sharp peak near the Fermi level ($E_{\textrm{F}}$).
A decrease of Co concentration leads to the depopulation of the valence band, 
whereby the relative position of the Fermi level moves towards the center of that peak.
Due to the exchange interactions that peak splits asymmetrically below the critical concentration ($x_{crit} \sim 0.925$).
The majority spin channel (the occupied one, red in Fig.~\ref{fig:dos}) moves towards lower energies by about 0.4~eV,
while the minority spin channel (the unoccupied one, blue) is shifted towards higher energies by about 0.2~eV.
Since the sharp peak of minority spin channel is located on the Fermi level, the DOS at Fermi level increases.
Further decrease in Co concentration $x$ leads to an increase of the magnetic moment and thus to increase of the exchange splitting. 
That shifts the sharp peak observed below the Fermi level towards the higher energies and the DOS at Fermi level decreases.

\begin{figure}[ht]
\includegraphics[trim = 110 35 10 20,clip,height=1.0\columnwidth,angle=270]{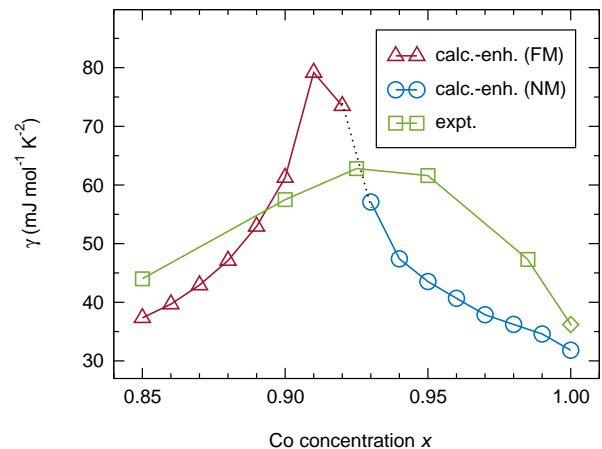}
\caption
{
\label{fig:gamma_exp_theory} 
The calculated (with LDA-VCA) enhanced electronic specific heat coefficient $\gamma_{calc-enh}$ for the \yfeco{} system in the Co-rich region ($0.85 \leq x < 1$), 
together with the experimental dependence of $\gamma_{\textrm{expt}}$.
The enhancement factor $\tilde{\gamma}$~=~6.87 was applied as calculated for \yco{} by Tanaka and Harima~\cite{tanaka_mass_1998}.
The $\gamma_{calc-enh}$ plot consists of two sections, ferromagnetic (FM) and nonmagnetic (NM), on the two sides of the phase transition determined by $x_{crit}\sim0.925$.
The value of $\gamma$ for \yco{} ($x=1.0$) comes from Muraoka et al.~\cite{muraoka_electronic_1977}.
}
\end{figure}

%---------------------gamma------------------------------
%
In the next step we calculate the values of $\gamma$ from DOS$(E_{\textrm{F}})$ according to Eq.~\ref{eq1}.
For the nonmagnetic \yco{} ($x = 1.00$) the $\gamma_{\textrm{calc}}$ is equal to 4.6~mJ\,mol$^{-1}$\,K$^{-2}$ 
and it is relatively close to a previously calculated value equal to 6.1~mJ\,mol$^{-1}$\,K$^{-2}$~\cite{tanaka_mass_1998}.
Both those calculated values are distant from the experimental one, equal to 36.2~mJ\,mol$^{-1}$\,K$^{-2}$ (per mol of \yco{})~\cite{muraoka_electronic_1977}.
The underestimation of $\gamma$ values is a recognized DFT weakness 
related to disregard of spin fluctuations and many-body effects in low-energy excitations~\cite{tanaka_mass_1998}.
The many-body effects can be taken into account by considering the self-energy of the correlated electrons.
By using that approach Tanaka et~al. justified the introduction of the so called enhancement factor ($\tilde{\gamma}$)~\cite{tanaka_calculations_1998, tanaka_mass_1998}, 
see Sec.~\ref{exp_comp_details} for more details.
The enhancement factor for \yco{}, obtained by Tanaka and Harima by adjusting the calculated values of specific heat coefficient to experimental value, is equal 6.87~\cite{tanaka_mass_1998}.
We use this single value to calculate the enhanced specific heat coefficient ($\gamma_{calc-enh}$) for each considered composition simply by multiplying the calculated with LDA specific heat coefficient $\gamma_{\textrm{calc}}$ by the enhancement factor $\tilde{\gamma} = 6.87$.
%
%In a result, the calculated values of $\gamma_{calc-enh}$ are in the same range as the values of $\gamma_{expt}$, see Fig.~\ref{fig:gamma_exp_theory}.
%
The $\gamma_{calc-enh}(x)$ dependence consists of two regions (ferromagnetic and nonmagnetic) separated by a phase transition at $x_{crit} \sim 0.925$.
Similar like the experimental result it shows a maximum for the Co-rich compositions close to  $x_{crit}$.
The calculated and experimental Co concentrations for which the maximum in $\gamma$ is observed are in good agreement with each other ($x_{max-LDA-VCA} = 0.91$ and $x_{max-expt} \sim 0.925$).
The shape of the $\gamma_{calc-enh}(x)$ dependence results directly from the contour of a narrow peak observed about 0.1 eV below $E_\textrm{F}$ in DOS plot of \yco{}, see Fig.~\ref{fig:dos}.
Position of this peak in relation to $E_\textrm{F}$ changes with $x$ and 
for the Co concentration range between about 0.85 and 1.00 this peak is fully \textit{scanned by the $E_\textrm{F}$}
which is reflected in the observed maximum in DOS($E_\textrm{F}$)($x$) and further in the corresponding $\gamma_{calc-enh}(x)$ dependence.

%------- Can the authors also clarify the connection between gamma_max, gamma_tilde=6.87, gamma_max_enh and gamma_max_expt, so readers who are not so familiar, can also gratify this paper.
%
The calculated (without enhancement) maximum of specific heat coefficient $\gamma_{max}$ is around 12~mJ\,mol$^{-1}$\,K$^{-2}$ (at $x_{max} = 0.91$).
This value multiplied by the enhancement parameter ($\tilde{\gamma}$~=~6.87) gives the maximum of enhanced specific heat coefficient $\gamma_{max-enh}\sim70$~mJ\,mol$^{-1}$\,K$^{-2}$,
whereas the experimental value $\gamma_{max-expt}$ is around 62.8~mJ\,mol$^{-1}$\,K$^{-2}$ (per mol of Y(Fe$_{0.075}$Co$_{0.925}$)$_2$) for $x=0.925$, see Table.~\ref{tab:gamma}.
The cause of big difference between the $\gamma_{max}$ and $\gamma_{max-expt}$, besides the disregard of many-body effects, 
can be due to not taking into account in theoretical models the additional impact of spin fluctuations around the magnetic phase transition~\cite{muraoka_electronic_1977}.
The enhanced specific heat coefficient $\gamma_{max-enh}$ is much closer to $\gamma_{max-expt}$, but at the price of using the enhancement parameter $\tilde{\gamma}$ justified by the model considering many-body effects but still neglecting spin fluctuations~\cite{tanaka_calculations_1998,tanaka_mass_1998}.

%-------------------------------small shift in position of the peak------------------------
%
It is a little bit surprising, that the calculated peak of $\gamma$ does not coincide with the $x_{crit-LDA-VCA} \sim 0.925$, but instead occurs at $x_{max} = 0.91$.
Unfortunately, the results of our measurements ($x_{max} \sim 0.925$ and $0.90 < x_{crit} < 0.95$) are not accurate enough to confirm this effect in \yfeco{}.
However, a much larger difference has been previously measured for Zr(Fe$_{1-x}$Co$_x$)$_2$ system, with $x_{max}=0.5$ and $x_{crit}=0.75$~\cite{hilscher_estimation_1978}.

%-------------theoretical sources of discrepancies 
%
The observed discrepancies between the experimental and computational results of $\gamma(x)$ 
arising, among others, from not considering many-body effects and spin fluctuations in the theoretical description
can be partly attributed also to the shortcomings of the LDA and VCA.
The application of LDA results in reduction of lattice parameter due to overbinding and it is accompanied by reduction of the magnetic moments, underestimation of the magnetic energy, and shift of the critical Co concentration for which the magnetic transition occurs.
Whereas, the VCA simplifies the nature of chemical disorder by forming a homogeneous crystal.
In result, we received a sharp band structure without any broadening coming from the chemical disorder, 
which is observed in the angle-resolved photoemission spectroscopy (ARPES) measurements and in the CPA calculations of disordered alloys.
So why did we choose the VCA instead CPA to do the research?
An application of CPA would not allow us to analyze the fixed spin moment and Fermi surface, which results will be presented in the following sections.
Nevertheless, we have also performed the additional CPA-LDA calculations, which have confirmed the presence of the magnetic phase transition and the maximum in $\gamma(x)$ for the Co-rich compositions.
However, the CPA maximum in $\gamma(x)$ is much smoother than that from VCA method, which can be related to the mentioned broadening caused by chemical disorder.
Some results of the CPA calculations will be also presented in the further part of the study.

\subsubsection{Fixed spin moment calculations}
%
%--------------------------dosEF vs fsm---------------------
%
\begin{figure}[!ht]
\includegraphics[trim = 20 0 10 10,clip,width=\columnwidth,angle=0]{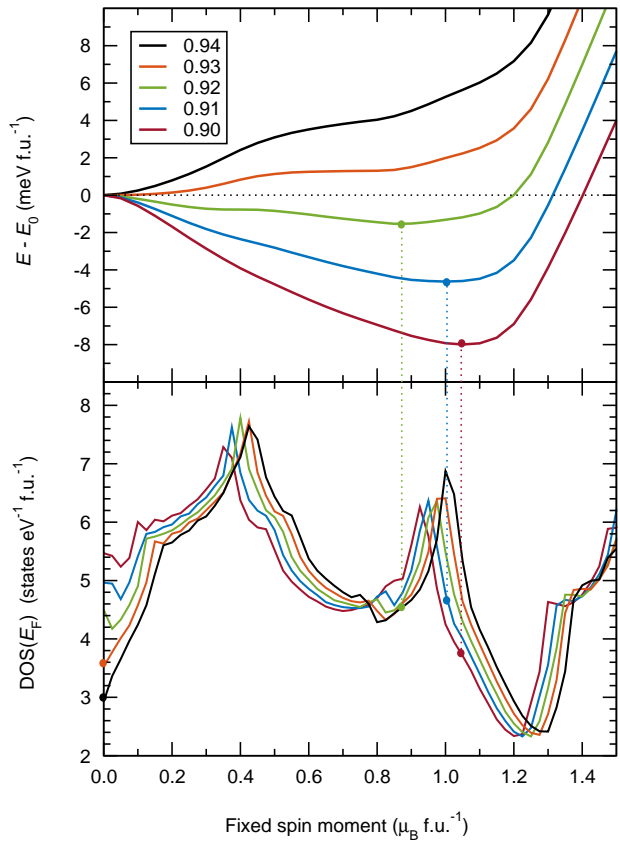}
\caption
{
\label{fig:dosEF_vs_fsm} 
The calculated (with LDA-VCA method) magnetic energy and density of states at Fermi level \textit{versus} fixed spin moment for several Co concentrations in the vicinity of the critical concentration $x_{crit} \sim 0.925$ at which occurs a ferromagnetic--nonmagnetic phase transition for the \yfeco{} system. 
The ground state DOS($E_{\textrm{F}}$) are marked with colored dots. 
The vertical dashed lines connect the magnetic energy minima with corresponding DOS($E_{\textrm{F}}$).
}
\end{figure}

The fixed spin moment (FSM) method allows for calculations of the systems with a non-equilibrium magnetic moment, 
and hence it enables to plot the magnetic energy dependence as a function of spin magnetic moment~\cite{schwarz_itinerant_1984},
where the magnetic energy is the energy difference between the magnetic and nonmagnetic ground state solutions.
The FSM calculations have already helped to understand the magnetic behavior of \yco{}~\cite{schwarz_itinerant_1984,rusz_magnetism_2004}.
We performed a series of FSM calculations for several successive concentrations near the magnetic transition of \yfeco{} system, see Fig.~\ref{fig:dosEF_vs_fsm}.
The presented dependences of magnetic energy and the density of states at Fermi level (DOS($E_{\textrm{F}}$)) on the FSM 
are intended to explain the behavior of $\gamma(x)$.
In order to obtain accurate plots of DOS($E_{\textrm{F}}$) \textit{versus} FSM, it was necessary to use a very small step in FSM (0.025~$\mu_{\textrm{B}}$).
The magnetic energy plots, on top panel, confirms that within the LDA-VCA the magnetic phase transition in \yfeco{} occurs at Co concentration equal to about 0.925.
The plots show minima for non-zero moments above this value.
The positions of the minima shift towards higher magnetic moments with a decrease of Co concentration,
wherein, the deeper is the minimum, the more stable the ferromagnetic state is.
The overall shape of DOS($E_{\textrm{F}}$)($m$) is similar for all considered Co concentrations.
The observed double peak structure is related to the DOS plot with the characteristic maximum near the Fermi level.
The observed for increasing Co concentration $x$ shift of the DOS($E_{\textrm{F}}$)($m$) plots towards the higher magnetic moments comes from filling of the electronic band structure of \yfeco{} after alloying with the element possessing more valence electrons.
The minima in magnetic energies clearly correspond with the ground state values of DOS($E_{\textrm{F}}$).
When we look at the values of the ground state DOS($E_{\textrm{F}}$) for subsequent Co concentrations, 
we observe the change of tendencies, 
leading to the formation of the maximum in DOS($E_{\textrm{F}}$) \textit{versus} $x$, 
and thus in the considered dependence of $\gamma(x)$.
It gives another perspective for understanding of the formation of maximum in $\gamma(x)$.

\subsubsection{Fermi surface}
%--------------------------Fermi surfaces---------------------
%
\begin{figure}[t]
\includegraphics[width=0.9\columnwidth]{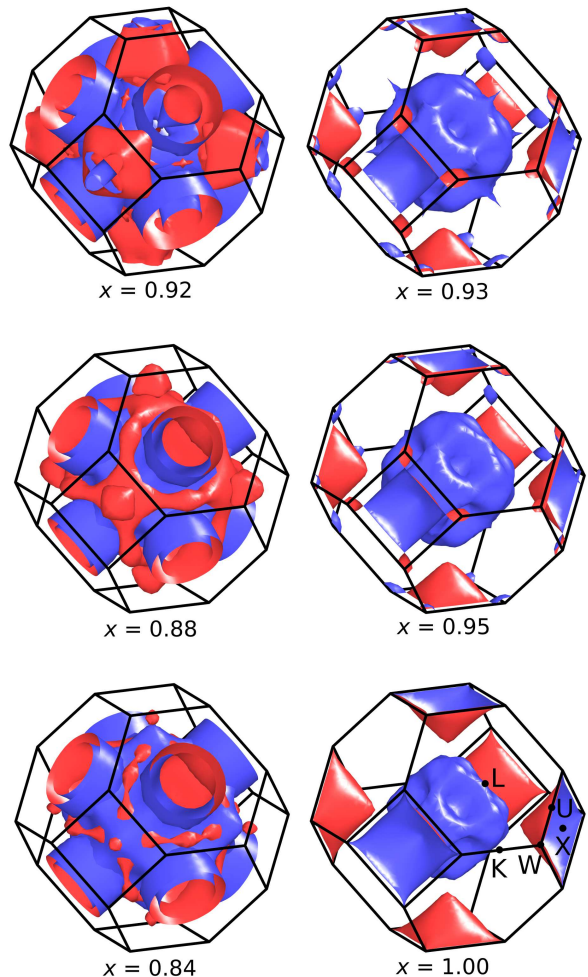}
\caption
{
\label{fig:fs} 
The Fermi surfaces of \yfeco{} calculated within LDA-VCA in the vicinity of the critical concentration $x_{crit} \sim 0.925$ at which the ferromagnetic--nonmagnetic phase transition occurs.
The results of spin polarized calculations are presented in the left column, 
while the results of nonmagnetic calculations are shown in the right one. 
Red color denotes the electron-type, while blue color the hole-type surfaces.
}
\end{figure}
%
%-----------------FS very general intro----------------------------
%
Another insight into the nature of  the electronic specific heat coefficient $\gamma$ can be achieved from a perspective of the Fermi surface.
As we discussed in Sec.~\ref{sec:intro}, $\gamma$ is directly estimated from DOS($E_{\textrm{F}}$), 
where the latter one is obtained by integration of the states on the Fermi level over the whole Brillouin zone.
A distribution of the states at the Fermi level in the first Brillouin zone is called the Fermi surface.
Thus, the observed maximum in $\gamma$ correlates with the largest area of the Fermi surface, as presented in Fig.~\ref{fig:fs} for  several successive Co concentrations.
The Fermi surfaces presented in the right column are nonmagnetic solutions,
while the ones shown in the left column are spin polarized.

%--------------fermi surface of yco2-------------------------------------
%
The nonmagnetic Fermi surface for the terminal concentration $x$~=~1.00 (\yco{}) consists of three sheets.
Electron-type surface, denoted with red color, is centered at high-symmetry point X and consists of rectangular parts connected at W points. 
The other two surfaces are of the hole-type and are located around the $\Gamma$ point.
One of them is nested and thus not visible in the figure.
The Fermi surface calculated for \yco{} is consistent with the previous theoretical results~\cite{tanaka_calculations_1998}.
We suspect that the small differences observed around the W point may come from the various forms of the LDA exchange-correlation potential used in the compared models 
or from the inclusion of the spin-orbit interactions in the previous model~\cite{tanaka_calculations_1998}.

%--------------------FS evolution----------------------------
%
An evolution of Fermi surface is observed with decrease of Co concentration ($x$~=~0.95 and 0.93).
The new features are forming around the W point.
The hole-type surfaces (blue color) are enlarging,
which leads to increase of DOS($E_{\textrm{F}}$) and eventually to fulfillment of the Stoner criterion at $x_{crit}\approx0.925$.
Decrease of $x$ leads to a magnetic phase transition, which also manifests in the shape of the Fermi surfaces.
The spin polarized Fermi surface for $x$~=~0.92 consists of two overlapping spin channels and has the largest area of all Fermi surfaces shown in Fig.~\ref{fig:fs}.
The characteristic feature of the spin polarized solutions are the hole-type nested double tubes along the $\Gamma$-L direction.
The Fermi surfaces for even lower Co concentrations ($x$~=~0.88 and 0.84) exhibit a gradual decrease of their area, 
which correlates with the form of the DOS($E_{\textrm{F}}$) and $\gamma(x)$ dependences.

\subsubsection{Magnetic moments from CPA and ordered compound method}

As an addition to the presented VCA results obtained for the Co-rich concentrations of the \yfeco{} system, 
we show the magnetic moments calculated using the coherent potential approximation (CPA) 
and ordered compound method for a full Co concentration range.

%--------------------magnetic moments----------------------------------
%
%top left bottom
\begin{figure}[t]
\includegraphics[trim = 60 55 10 60,clip,height=1\columnwidth,angle=270]{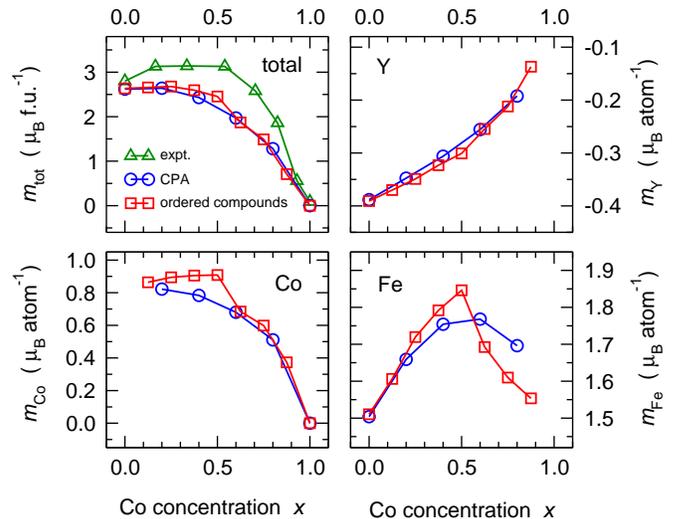}
\caption
{
\label{fig:mom} 
The spin magnetic moments \textit{versus} Co concentration $x$ for the \yfeco{} system as calculated with the FPLO-LDA.
For YFe$_2$ and YCo$_2$ models we used optimized lattice parameters 
and for the intermediate concentrations the lattice parameters obtained using the linear interpolation.
The subsequent panels present total magnetic moments and contributions from individual elements.
The chemical disorder was treated with the coherent potential approximation (CPA) and the ordered compound's method.
The experimental total magnetic moments' dependence comes from the work of Piercy and Taylor~\cite{piercy_evidence_1968}.
}
\end{figure}
The magnetic moments calculated with the CPA and ordered compound methods are shown in Fig.~\ref{fig:mom}.
%
%The results of total magnetic moment preserve the overall shape of the Slater-Pauling curve with a maximum at $x\sim0.3$.
%
%--------difference between magnetic moments from calculations and experiment - limits
%
The calculated total magnetic moments are underestimated in comparison to the experimental values~\cite{piercy_evidence_1968}.
This difference is relatively large and for the intermediate concentrations it is equal to about 0.7~$\mu_{\textrm{B}}$\,f.u.$^{-1}$.
That discrepancy originates, among others, from the limitations of the LDA, which is recognized as an underestimating of magnetic moment~\cite{werwinski_magnetocrystalline_2018}.
Simultaneously, we used the lattice parameters calculated within the LDA (underestimated by about 0.3~\AA{})~\cite{guzdek_electrical_2012}, which additionally decreased the magnetic moment.
The application of generalized gradient approximation (GGA) gave better values of optimized lattice parameters and magnetic moments, but it overestimated the magnetic energy,
leading to a ferromagnetic ground state for \yco{} and no magnetic phase transition in \yfeco{} was observed~\cite{wasilewski_curie_2018}.
Furthermore, the lattice parameters for the intermediate \yfeco{} concentrations were interpolated assuming a linear behavior of the $a(x)$ dependence, which further decreased the lattice parameters of the intermediate concentrations in respect to the experimental values~\cite{guzdek_electrical_2012}. 
Some effect on the calculated magnetic moments has an application of the scalar-relativistic approximation,  
as a result of which the obtained magnetic moments are completely spin type and do not include the orbital contributions.
This deficiency is the magnitude of the orbital moments of the bcc Fe and hcp Co, which have experimental values equal to 0.086 and 0.13~$\mu_{\textrm{B}}$, respectively~\cite{chen_experimental_1995,moon_distribution_1964}.
Previously observed failure of the LDA~+~U approach for \yco{} suggests a necessity to make use of dynamical correlations to improve the theoretical model~\cite{wasilewski_curie_2018}.
However, this goes beyond the scope of this work.

The calculated total magnetic moment for \yfe{}, equal to 2.62~$\mu_{\textrm{B}}$\,f.u.$^{-1}$, is lower than the experimental value of 2.80~$\mu_{\textrm{B}}$\,f.u.$^{-1}$~\cite{piercy_evidence_1968}.
The contributions to the total magnetic moment from Fe and Y atoms are equal to 1.51~$\mu_{\textrm{B}}$ and -0.39~$\mu_{\textrm{B}}$, respectively, in qualitative agreement with the previous theoretical results (1.68~$\mu_{\textrm{B}}$ and -0.43~$\mu_{\textrm{B}}$~\cite{eriksson_electronic_1988}).
The calculated spin magnetic moment on Fe in \yfe{} (1.51~$\mu_{\textrm{B}}$) is reduced in comparison to the measured spin magnetic moment of bcc-Fe, equal to 1.98~$\mu_{\textrm{B}}$~\cite{chen_experimental_1995}.
Similarly, the highest values of magnetic moment on Co in the \yfeco{} system, equal to about 0.8~$\mu_{\textrm{B}}$ on the Fe-rich limit, are significantly reduced in respect to the experimental magnetic moment of fcc-Co, equal to 1.67~$\mu_{\textrm{B}}$~\cite{reck_orbital_1969}.
The above picture of the magnetic properties obtained from the CPA and ordered compound method supports the previous findings based on VCA and helps to better understand the behavior of experimentally observed $\gamma$ \textit{versus} $x$ dependence.

\section{Summary and conclusions}
In this work we have presented a combined experimental and computational study of the electronic specific heat coefficient $\gamma$ and magnetic properties of \yfeco{} Laves phases.
For high Co concentrations the measurements indicated the
presence of the concentration induced ferromagnetic--paramagnetic phase transition accompanied with the maximum in $\gamma$.
The magnetic transition was modeled based on the LDA ground state electronic structure.
Calculations showed also that the observed maximum in $\gamma(x)$ results from the presence of the sharp DOS peak on the Fermi level.
The calculated with LDA values of $\gamma$ are significantly underestimated because of not including the many-body effects and spin fluctuations in the LDA description.
To improve the calculated $\gamma$ values we used the so called enhancement factor, introduced and evaluated by another group.
The introduction of the enhancement factor lets to incorporate the many-body effects by considering the self-energy of the correlated electrons.
In this work another perspective for understanding the formation of maximum in $\gamma(x)$
was given by simultaneous analysis of magnetic energy and DOS at the Fermi level as the functions of fixed spin moment.
Furthermore, using the CPA and ordered compound methods, we calculated the basic magnetic properties of \yfeco{} in the whole range of concentrations.

\begin{acknowledgments}
B.W. and M.W. acknowledge the financial support from the Foundation of Polish Science grant HOMING.
The HOMING programme is co-financed by the European Union under the European Regional Development Fund.
M.W. acknowledges the financial support of the National Science Centre Poland under the decision DEC- 2018/30/E/ST3/00267.
J.R. acknowledges the financial support from Swedish Research Council.
O.E. acknowledges support from the Swedish Research Council, STandUPP, eSSENCE and the KAW foundation.
Part of the computations was performed on the resources provided by the Poznań Supercomputing and Networking Center (PSNC).
\end{acknowledgments}

\appendix*
\section{Crystallographic data of ordered compounds}
\begin{table}[!ht]
\caption{\label{tab:ordered_compounds} 
Crystallographic data for Y$_4$Fe$_7$Co$_1$, Y$_4$Fe$_6$Co$_2$, Y$_4$Fe$_5$Co$_3$, and Y$_4$Fe$_4$Co$_4$ ordered compounds, where $a$ is the cubic Laves phase lattice parameter (optimized $a$ equals 7.045~\AA{} for \yfe{} and 6.95~\AA{} for \yco{}). }
\begin{tabular}{|cc|ccc||cc|ccc|}
\hline \hline
\multicolumn{5}{|c||}{Y$_4$Fe$_7$Co$_1$ sg. 20 $C222_1$} & \multicolumn{5}{c|}{Y$_4$Fe$_5$Co$_3$ sg. 20 $C222_1$}   \\
\hline
	&$a,b,c$& $a\sqrt{2}$	& $a\sqrt{2}$ 	& $a$ 	&  	&$a,b,c$& $a\sqrt{2}$	& $a\sqrt{2}$	& $a$    \\
	& atom  &  $x$ 		& $y$ 		& $z$ 	&     	& atom 	&  $x$ 		& $y$ 		& $z$ \\
\hline
 1  	&  Y    &     1/8  	& 1/8  		& 1/8 	&  1  	&  Y    &     3/8  	& 7/8  		& 1/8 \\ 
 2  	&  Y    &     1/8  	& 3/8  		& 7/8 	&  2  	&  Y    &     7/8  	& 1/8  		& 7/8 \\
 3  	&  Fe   &     1/2  	& 1/8  		& 3/4 	&  3  	&  Fe   &     1/2  	& 3/8  		& 3/4 \\
 4  	&  Fe   &     3/8  	& 1/4  		& 0   	&  4  	&  Fe   &     5/8  	& 1/4  		& 0   \\ 
 5  	&  Fe   &     3/4  	& 1/8  		& 3/4 	&  5  	&  Fe   &     1/4  	& 3/8  		& 3/4 \\ 
 6  	&  Fe   &     5/8  	& 0    		& 1/2 	&  6  	&  Co   &     7/8  	& 0    		& 1/2 \\ 
 7  	&  Fe   &     1/2  	& 3/8  		& 1/4 	&  7  	&  Co   &     1/2  	& 1/8  		& 1/4 \\ 
 8  	&  Co   &     1/8  	& 0    		& 1/2 	&  8  	&  Co   &     3/8  	& 0    		& 1/2 \\ 
\hline 
 \multicolumn{5}{|c||}{Y$_4$Fe$_6$Co$_2$ sg. 213 $P4_{1}32$} & \multicolumn{5}{c|}{Y$_4$Fe$_4$Co$_4$ sg. 91 $P4_{1}22$}  \\
\hline
	&$a,b,c$& $a$ 		& $a$ 		& $a$ 	& 	&$a,b,c$& $a/\sqrt{2}$ 	& $a/\sqrt{2}$ 	& $a$\\
	&atom   &  $x$ 		& $y$ 		& $z$  	&  	& atom  &  $x$ 		& $y$ 		& $z$ \\
\hline
 1  	&  Y    &     0    	& 1/2  		& 1/2 	&  1  	&  Y    &     3/4  	& 3/4  		& 7/8 \\ 
 2  	&  Co   &     3/8  	& 3/8  		& 3/8 	&  2  	&  Fe   &     1/4  	& 0    		& 3/4 \\ 
 3  	&  Fe   &     7/8  	& 7/8  		& 3/8 	&  3  	&  Co   &     3/4  	& 1/2  		& 1/4 \\
\hline \hline
\end{tabular}
\end{table}

One of the common approximate techniques for a computational treatment of the structures with a chemical disorder is the ordered compound method~\cite{eriksson_electronic_1989}.
It assumes simulating of a disordered alloy by the ordered compound of the same composition.
For example, in case of the \yfeco{} alloys one can represent Co concentration $x=0.125$ with an ordered ternary compound Y$_4$Fe$_7$Co$_1$.
The generated structural data of the ordered ternary compounds for $x$ equal 0.125, 0.25, 0.375, and 0.5 are presented in Table~\ref{tab:ordered_compounds}. 
The other three intermediate compositions, 0.625, 0.75, and 0.875, can be obtained from the presented data by exchanging atoms on Fe and Co sites.
The only free parameter here is the lattice parameter $a$, thus these ordered compounds are universal and can be used for calculations of alloys of any AB$_2$-type Laves phase.

\newpage
\end{sloppypar}

\bibliography{yfeco2}

\end{document}